\documentclass[final,5p,times,twocolumn]{elsarticle}

\usepackage{amssymb}
\usepackage{siunitx}
\usepackage{enumitem}

\usepackage{amsfonts,amssymb,amsmath,amsbsy}
\usepackage{physics}
\usepackage{color,calc}
\usepackage{bm}
\usepackage{array}
\usepackage{booktabs}
\usepackage{graphicx}
\usepackage{graphics}
\usepackage{eepic,epsfig}
\usepackage[breaklinks,hidelinks,colorlinks=true,linkcolor=blue,citecolor=blue,filecolor=blue,urlcolor=blue]{hyperref}

\usepackage[dvipsnames]{xcolor}
\definecolor{myblue}{RGB}{0, 0, 255}
\usepackage[toc,page]{appendix}

\usepackage{microtype}

\usepackage[dvipsnames]{xcolor}
\definecolor{myblue}{RGB}{0, 0, 255}
\usepackage[toc,page]{appendix}

\def\be{ \begin{equation} }
\def\ee{ \end{equation} }
\def\bea{ \begin{eqnarray} }
\def\eea{ \end{eqnarray} }
\def\bse{ \begin{subequations} }
\def\ese{ \end{subequations} }
\def\ba{ \begin{array} }
\def\ea{ \end{array} }
\def\bt{ \begin{tabular} }
\def\et{ \end{tabular} }

\def\etal{\textit{et al.}}

\long\def\/*#1*/{}

\newcommand{\rom}[1]{\expandafter\romannumeral #1\relax}

\journal{Radiation Physics and Chemistry}

\begin{document}

\begin{frontmatter}



\title{Positron Bunch Radiation in the System of Tightly Packed Nanotubes}


\author[inst1,inst2]{Hayk L. Gevorgyan}

\cortext[cor1]{Corresponding author}
\ead{hayk.gevorgyan@aanl.am}
\affiliation[inst1]{organization={Experimental Physics Division, A.I. Alikhanyan National Science Laboratory (Yerevan Physics Institute)},%Department and Organization
            addressline={2 Alikhanyan Brothers St.}, 
            city={Yerevan},
            postcode={0036}, 
            country={Armenia}}
            
\affiliation[inst2]{organization={Quantum Technologies Division, A.I. Alikhanyan National Science Laboratory (Yerevan Physics Institute)},%Department and Organization
            addressline={2 Alikhanyan Brothers St.}, 
            city={Yerevan},
            postcode={0036}, 
            country={Armenia}}

\author[inst3]{Koryun L. Gevorgyan}

\affiliation[inst3]{organization={Department of Physics, Yerevan State University},%Department and Organization
            addressline={1 Alex Manoogian St.}, 
            city={Yerevan},
            postcode={0025}, 
            country={Armenia}}

\author[inst4]{Lekdar A. Gevorgian}

\affiliation[inst4]{organization={Matinyan Center for Theoretical Physics, A.I. Alikhanyan National Science Laboratory (Yerevan Physics Institute)},%Department and Organization
            addressline={2 Alikhanyan Brothers St.}, 
            city={Yerevan},
            postcode={0036}, 
            country={Armenia}}

\begin{abstract}
Radiation emitted by a bunch of positrons channeled in nanotubes at zero emission angle is studied taking into account medium polarization. The formation of radiation is characterized by an energy threshold that depends on the oscillation amplitude of each positron. When the bunch energy reaches the maximum value of the threshold energy, radiation is produced by all positrons in the bunch. The nanotube potential barrier is described using a harmonic model. The spectral line shape of the radiation from the positron bunch, the fundamental radiation frequency, and the number of emitted photons are determined. It is shown that a system of tightly packed carbon nanotubes can generate an intense, quasi-monochromatic, and directed beam of circularly polarized soft X-ray photons with an energy of about $3$~\SI{}{\kilo\electronvolt} (wavelength $4.1$~\SI{}{\angstrom}).

% The problem of channeling radiation of positron bunch in the system of packed nanotubes was investigated in the present work. Used the model of harmonic potential which is justified since on the one hand the number of positrons in the region near the axis of nanotube is small, and on the other hand their contribution to the formation of the total radiation is also small. The problem is solved in the dipole approximation. The radiation at first harmonic occurs at zero angle too. At zero angle are radiated both extremely hard and extremely soft photons due to the medium polarization. The frequency-angular distribution of number of emitted photons was received. The distribution does not depend on the azimuthal angle, since the task has cylindrical symmetry. Radiation at the zero angle is fully circularly polarized. For formation of radiation there is an energy threshold, which depends on the oscillation amplitude of channelling positrons. When the bunch energy coincides with the upper threshold then in radiation contribute all channeled positrons. Thus is formed intensive, quasi-monochromatic and circularly polarized X-ray photon beam which may have important practical application.
\end{abstract}


% \begin{highlights}
% \end{highlights}


\end{frontmatter}



%%%%%%%%%%%%%%%%%%%%%%%%%%%%%%%%%%%%%%%%%%%%%%%%%%%%%%%%%%%%%%%%%%%%%%%%%%%%%%%%%%%%%%%%%%%%%%%%%%%%%%%%%%%%%%%%%%%%%%%%%%%%%%%%%%%%%%%%%
%%%%%%%%%%%%%%%%%%%%%%%%%%%%%%%%%%%%%%%%%%%%%%%%%%%%%%%%%%%%%%%%%%%%%%%%%%%%%%%%%%%%%%%%%%%%%%%%%%%%%%%%%%%%%%%%%%%%%%%%%%%%%%%%%%%%%%%%%
\section{Introduction}
%%%%%%%%%%%%%%%%%%%%%%%%%%%%%%%%%%%%%%%%%%%%%%%%%%%%%%%%%%%%%%%%%%%%%%%%%%%%%%%%%%%%%%%%%%%%%%%%%%%%%%%%%%%%%%%%%%%%%%%%%%%%%%%%%%%%%%%%%
%%%%%%%%%%%%%%%%%%%%%%%%%%%%%%%%%%%%%%%%%%%%%%%%%%%%%%%%%%%%%%%%%%%%%%%%%%%%%%%%%%%%%%%%%%%%%%%%%%%%%%%%%%%%%%%%%%%%%%%%%%%%%%%%%%%%%%%%%

In 1947, V. L. Ginzburg proposed the idea of generating electromagnetic radiation in the submillimeter range using relativistic electrons oscillating in periodic electric fields \cite{ginzburg1947}. Motz developed the theory of radiation of relativistic electrons passing through periodic field generated by magnets (undulator) \cite{motz1951}. Under the leadership of Motz, experimental studies were conducted to detect the radiation from relativistic electrons passing through a magnetic undulator \cite{motz1953}. It was found high intensity radiation in the millimeter range. Such intensity was obtained due to coherent radiation of a bunch the longitudinal size of which is smaller than the radiation wavelength. As was shown in \cite{korkhmazian1977} for shorter wavelengths regardless of the radiation type the coherence factor can exceed unity when the electron distribution is asymmetrical in the longitudinal direction. The effect of partially coherent radiation of bunch was detected in the experimental work \cite{ishi1995}. The parially coherent radiation of asymmetrical bunch will increase the efficiency of free electron laser (FEL) \cite{gevorgian1982}. The gain of FEL as has been shown by Madey depends on the line shape of the spontaneous emission \cite{deacon1977, madey1977}. Spontaneous undulator radiation in the X-ray frequency range was investigated by Korkhmazyan \cite{korkhmazian1970, korkhmazian1970aa}, and the experiments were carried out on Yerevan's accelerator \cite{alikhanyan1972, ispirian1971} to detect this radiation. In the formation of X-ray undulator radiation the medium polarization has a significant role \cite{gevorgyan1979}. The radiation is generated when the bunch energy is greater than the threshold energy. When the energy is close to the threshold, the frequency-angular distribution is narrowed, i.e., the emitted photon density increases \cite{gevorgian1977, gevorgian1977a, gevorgian1977b}. X-ray FEL was observed in the SASE FEL experiment \cite{emma2010}. Crystal can perform the role of peculiar microundulator for the channeled charged particles. As a result \cite{robinson1963, beeler1963} of numerical modeling of the process when fast electrons penetrate into monocrystal it has been observed that in certain crystal orientations mean free path of ions increases abnormally (channeling). The phenomenon of channeling has been observed experimentally in \cite{lindhard1965, piercy1963} and was explained by Lindhard in the work \cite{lutz1963} where the true potential of the crystal was replaced by the continuous potential averaged over atom coordinates. The theory of channeling radiation of charged particles has been developed by Kumakhov \cite{kumakhov1976}. This topic has been the subject of numerous theoretical and experimental studies \cite{bazylev1987}. The oscillation frequency of channeled particles in the crystalline or nanotube undulators with the harmonic potential depends on the particle energy. In the case of a nonharmonic potential, this frequency also depends on the oscillation amplitude \cite{gevorgian1998}. In periodically bent crystals, in addition to channeling radiation, undulator radiation is also produced due to the periodicity of the particles’ average trajectory \cite{kaplin1980}. The characteristics of the particle radiation, generated in a crystalline undulator, were investigated in \cite{korol1999}. X-ray and neutron channeling in carbon nanotubes is studied in \cite{dabagov2011}. The spontaneous and stimulated radiation in the crystalline or nanotube undulators has been studied taking into account the medium polarization \cite{avakyan1998, avakyan2001}. Due to the centrifugal force \cite{tsyganov1976, tsyganov1976a}, the dechanneling of positrons does not occur if the maximum bending angle of the crystalline undulator is smaller than the Lindhard angle \cite{gevorgian2005}. Particle acceleration in crystalline and nanotube undulators, taking into account medium polarization, is studied in \cite{wiedemann2001}. Recent studies have further investigated coherent radiation characteristics of modulated electron and positron bunches, X-ray crystalline undulator radiation, and the influence of dispersive media and FEL modulation on radiation gain and line shape \cite{HGevorgyan2017, HGevorgyan2021a, HGevorgyan2021b, Shamamian2024, HGevorgyan2024a, HGevorgyan2024b, HGevorgyan2025a}.

In this paper we derived the spectral distribution of total radiation of channeled positrons at normal incidence to the tightly-packed nanotube system.

%%%%%%%%%%%%%%%%%%%%%%%%%%%%%%%%%%%%%%%%%%%%%%%%%%%%%%%%%%%%%%%%%%%%%%%%%%%%%%%%%%%%%%%%%%%%%%%%%%%%%%%%%%%%%%%%%%%%%%%%%%%%%%%%%%%%%%%%%
%%%%%%%%%%%%%%%%%%%%%%%%%%%%%%%%%%%%%%%%%%%%%%%%%%%%%%%%%%%%%%%%%%%%%%%%%%%%%%%%%%%%%%%%%%%%%%%%%%%%%%%%%%%%%%%%%%%%%%%%%%%%%%%%%%%%%%%%%
\section{The trajectory of a channeled positron in the nanotube potential barrier}
%%%%%%%%%%%%%%%%%%%%%%%%%%%%%%%%%%%%%%%%%%%%%%%%%%%%%%%%%%%%%%%%%%%%%%%%%%%%%%%%%%%%%%%%%%%%%%%%%%%%%%%%%%%%%%%%%%%%%%%%%%%%%%%%%%%%%%%%%
%%%%%%%%%%%%%%%%%%%%%%%%%%%%%%%%%%%%%%%%%%%%%%%%%%%%%%%%%%%%%%%%%%%%%%%%%%%%%%%%%%%%%%%%%%%%%%%%%%%%%%%%%%%%%%%%%%%%%%%%%%%%%%%%%%%%%%%%%

% Let a positron bunch with a transverse uniform distribution move parallel to the axis of tightly-packed nanotubes. Then, approximately 90\% of the positrons are channeled inside the nanotube \cite{gevorgian1998}. 

% It has been shown that the potential of the form $V(s) = U_0 s^6$ is in good agreement with the calculated potential, where $U_0$ is the height of the potential barrier in energy units, $s=r/R$ ($0 \leq s \leq 1$), $r$ is initial distance from the axis of the nanotube, $R$ is the nanotube radius. 

% % In this work the trajectories of channeled positrons have been calculated.

% % We are interested in the total spectrum of the radiation intensity of all channeled positron of bunch.

% % The choosing of potential type $V(s)$ conditioned by that it also agrees with calculated potential also for the small values of $s$. However, in this case, on the one hand it is complicate calculation of the spectral intensity and on the other hand is lost the monochromaticity of radiation.

% % Therefore, the contribution of positrons with small $s$ to the total spectrum can be neglected, since both their number and emission intensity ($\sim s^2$) are small.

In this paper, to find the frequency–angular distribution of the radiation intensity of channeled positrons, we use a harmonic potential model to avoid more complicated calculations,
\be\label{}
U(s) = U_0 s^2,
\ee
which differs significantly from other analytically solvable models ($\propto s^4$ and $\propto s^6$ \cite{gevorgian1998}) only for small values of $s$, whose contribution to the radiation is negligibly small.

In the such potential barrier of nanotube the positrons are oscillating with the same frequency $\Omega_{ch}$:
\be\label{}
\begin{gathered}
\Omega_{ch} = \frac{\Omega_0}{\sqrt{\gamma}}, \quad \Omega_0 = \frac{c \sqrt{2\nu}}{R} = \frac{2\pi c}{l_0}, \quad \nu = \frac{U_0}{m c^2}, \\
l = l_0 \sqrt{\gamma} = \frac{2\pi R}{\Theta_L}, \quad l_0 = \frac{2 \pi R}{\sqrt{2 \nu}}, \quad \Theta_L = \sqrt{\frac{2 \nu}{\gamma}},
\end{gathered}
\ee
where $\gamma$ is the relativistic Lorentz factor, $c$ is the speed of light, $m$ is the positron mass, $l$ is the spatial period of positron oscillations, $\Theta_L$ is the Lindhard channeling angle, and $R$ is the nanotube radius.

A channeled positron with an initial radial coordinate $s R$ ($0 < s \leq 1$) moves along a helical trajectory
\be
\vb*{r} (s, t) = \left\{s R \cos{\left(\Omega_{ch} t\right)} \, \vb*{e}_x, \quad s R \sin{\left(\Omega_{ch} t\right)} \, \vb*{e}_y, \quad \beta_z (s) c t \, \vb*{e}_z\right\},
\ee
with velocity (in units of the speed of light $c$ in vacuum)
\be
\vb*{\beta} (s, t) = \left\{ - \beta_\perp (s) \sin{\left(\Omega_{ch} t\right)} \, \vb*{e}_x, \quad \beta_\perp (s) \cos{\left(\Omega_{ch} t\right)} \, \vb*{e}_y, \quad \beta_z (s) \, \vb*{e}_z\right\},
\ee
where $\beta_\perp (s) = s \beta_\perp$, $\beta_\perp = \Omega_{ch} R / c = \sqrt{2 \nu / \gamma}$, $\vb*{e}_x$ and $\vb*{e}_y$ are unit vectors of the Cartesian coordinate system, and $\vb*{e}_z$ is the longitudinal unit vector directed along the nanotube axis.

Neglecting small losses due to radiation primarily caused by transverse motion, the particle's energy remains constant. For a given value of $s$
\be
\beta^2_z (s) = \beta^2 - \beta^2_\perp (s).
\ee
Taking into account $\beta^2_z (s) = 1 - \gamma^{-2}_z$, $\beta^2 (s) = 1 - \gamma^{-2}$, we have
\be
\gamma^2_z (s) = \gamma^2 / Q(s), \quad Q(s) = 1 + q^2 s^2, \quad q = \beta_\perp \gamma = \sqrt{2 \nu \gamma},
\ee
where $q$ is the radiation parameter for $s=1$. Hence, 
\be
\beta_z (s) = 1 - \frac{Q(s)}{2 \gamma^2}.
\ee

%%%%%%%%%%%%%%%%%%%%%%%%%%%%%%%%%%%%%%%%%%%%%%%%%%%%%%%%%%%%%%%%%%%%%%%%%%%%%%%%%%%%%%%%%%%%%%%%%%%%%%%%%%%%%%%%%%%%%%%%%%%%%%%%%%%%%%%%%
%%%%%%%%%%%%%%%%%%%%%%%%%%%%%%%%%%%%%%%%%%%%%%%%%%%%%%%%%%%%%%%%%%%%%%%%%%%%%%%%%%%%%%%%%%%%%%%%%%%%%%%%%%%%%%%%%%%%%%%%%%%%%%%%%%%%%%%%%
\section{Radiation field of a positron at zero emission angle with initial radial coordinate $s$}
%%%%%%%%%%%%%%%%%%%%%%%%%%%%%%%%%%%%%%%%%%%%%%%%%%%%%%%%%%%%%%%%%%%%%%%%%%%%%%%%%%%%%%%%%%%%%%%%%%%%%%%%%%%%%%%%%%%%%%%%%%%%%%%%%%%%%%%%%
%%%%%%%%%%%%%%%%%%%%%%%%%%%%%%%%%%%%%%%%%%%%%%%%%%%%%%%%%%%%%%%%%%%%%%%%%%%%%%%%%%%%%%%%%%%%%%%%%%%%%%%%%%%%%%%%%%%%%%%%%%%%%%%%%%%%%%%%%

As is well known, the radiation field of a charged particle is expressed as an integral along its trajectory, with integration performed over time. Since the transit time of a channeled positron through the nanotube depends on the parameter $s$, the integration must be carried out over the longitudinal coordinate $z$.

% $\Omega_0$ and $l_0$ are the natural frequency and the spatial period of nanotube.

The radiation field with frequency $\omega$, produced in a nanotube of length $L = n l_{ch}$ by a channeled positron with initial radial coordinate $s$ (where $n$ is the number of positron oscillations in the bunch), is given by

\be\label{radiationfield}
\vb*{E} (\omega, s) = \frac{1}{c \beta_z (s)} \int\limits_{-L/2}^{L/2} \vb*{a} (z,s) \exp{i b (z,s) z} \, dz,
\ee
where, taking into account the condition $\vb*{n} = \{0,0,\vb*{e}_z\}$ for the unit vector in the direction of zero-angle radiation, the vector $\vb*{a}(z,s)$ has the form
\be\label{parametera}
\begin{gathered}
\vb*{a} (z,s) = [\vb*{n} \times [\vb*{n} \times \vb*{\beta} (z, s)]] = \vb*{n} (\vb*{n} \vb*{\beta} (z,s)) - \vb*{\beta}(z,s) = \\
= - \vb*{\beta}_\perp (z,s) = \frac{s \beta_\perp}{2} \left((- \vb*{e}_y + i \vb*{e}_x) \exp{- i \frac{\Omega_{ch}}{c \beta_z (s)} z} - \right. \\
\left. - (\vb*{e}_y + i \vb*{e}_x) \exp{i \frac{\Omega_{ch}}{c \beta_z (s)} z} \right).
\end{gathered}
\ee
The argument of the exponential function is
\be
b(z,s) = \frac{\omega}{c \beta_z (s)} \left(1 - \beta_z (s) \sqrt{\varepsilon(\omega)} \right). 
\ee

Here the dielectric permittivity can be represented, for radiation frequencies $\omega$ much larger than the plasma frequency $\omega_p$, as
\be
\sqrt{\varepsilon(\omega)} = 1 - \frac{\omega^2_p}{2\omega^2}.
\ee

Since only the first term in expression \eqref{parametera} satisfies the energy-momentum conservation law for radiation emission, the integrand in \eqref{radiationfield}, taking into account \eqref{parametera} and $\beta_z (s) = 1 - \gamma^{-2}_z (s)/2$ can be written as
\be
\begin{gathered}
\frac{s \beta_\perp}{2} (- \vb*{e}_y + i \vb*{e}_x) \exp{i \frac{\omega}{2 c \beta_z (s) \gamma^2_z (s)} \left(1 - 2 \frac{\Omega_{ch} \gamma^2_z (s)}{\omega} + \frac{\omega^2_p \gamma^2_z (s)}{\omega^2} \right) z} = \\
= \frac{s \beta_\perp}{2} ( - \vb*{e}_y + i \vb*{e}_x) \exp{i \frac{\pi}{l_{ch} \beta_z (s) x} \varphi (x) z}, \\
\varphi (x) = x^2 - 2 x + \left(\frac{\omega_p}{\Omega_-}\right)^2 \frac{Q(s)}{\gamma},
\end{gathered}
\ee
where $x = \omega / (\Omega_{ch} \gamma^2_z (s))$ is the dimensionless frequency.

After integration, the radiation field takes the form
\be
\vb*{E} (x,s) = \frac{s \beta_\perp l_{ch}}{2 c \beta_z (s)} ( - \vb*{e}_y + i \vb*{e}_x) \frac{\sin{\left(n Y (x,s)\right)}}{Y(x,s)}, \quad Y(x,s) = \frac{\pi}{2} \frac{\varphi(x)}{\beta_z (s) x}.
\ee

%%%%%%%%%%%%%%%%%%%%%%%%%%%%%%%%%%%%%%%%%%%%%%%%%%%%%%%%%%%%%%%%%%%%%%%%%%%%%%%%%%%%%%%%%%%%%%%%%%%%%%%%%%%%%%%%%%%%%%%%%%%%%%%%%%%%%%%%%
%%%%%%%%%%%%%%%%%%%%%%%%%%%%%%%%%%%%%%%%%%%%%%%%%%%%%%%%%%%%%%%%%%%%%%%%%%%%%%%%%%%%%%%%%%%%%%%%%%%%%%%%%%%%%%%%%%%%%%%%%%%%%%%%%%%%%%%%%
\section{Spectral line shape of radiation from a positron channeled in a nanotube with oscillation amplitude $s$ ($0<s\leq 1$)}
%%%%%%%%%%%%%%%%%%%%%%%%%%%%%%%%%%%%%%%%%%%%%%%%%%%%%%%%%%%%%%%%%%%%%%%%%%%%%%%%%%%%%%%%%%%%%%%%%%%%%%%%%%%%%%%%%%%%%%%%%%%%%%%%%%%%%%%%%
%%%%%%%%%%%%%%%%%%%%%%%%%%%%%%%%%%%%%%%%%%%%%%%%%%%%%%%%%%%%%%%%%%%%%%%%%%%%%%%%%%%%%%%%%%%%%%%%%%%%%%%%%%%%%%%%%%%%%%%%%%%%%%%%%%%%%%%%%

For the frequency–angular distribution of the number of photons, we have \cite{jackson1999}
\be
\begin{gathered}
\frac{d^2 N(x,s)}{dx dO} = \frac{d^2 N(\omega, s)}{d\omega dO} \cdot \frac{d\omega}{dx} = \frac{\alpha \Omega_{ch} \gamma^2_z (s) x}{4 \pi^2} \Omega_{ch} \gamma^2_z (s) \abs{\vb*{E} (x,s)}^2, \\
\abs{\vb*{E} (x,s)}^2 = \frac{s^2 \beta^2_\perp l^2_{ch}}{2 c^2 \beta^2_z (s)} \frac{\sin^2 (n Y)}{Y^2}.
\end{gathered}
\ee

Here, $dO= \sin{\Theta} d\Theta d\varphi$ is the solid angle of radiation; $\theta = \vartheta \gamma_z (s)$ is the polar angle of radiation measured in units of $\gamma^{-1}_z(s)$ for relativistic particles, and $\varphi$ is the azimuthal angle.

For radiation at zero angle, we have $\int dO = 2\pi$.

Taking into account the equalities
\be
\begin{gathered}
l_{ch} \Omega_{ch} = 2 \pi c, \\
\gamma^2_z (s) = \frac{\gamma^2}{Q(s)} = \frac{\gamma^2}{1+q^2 s^2},\\
q^2 = (\beta_\perp \gamma)^2 = 2 \nu \gamma,
\end{gathered}
\ee 
where $q = \sqrt{2 \nu \gamma}$ is the radiation parameter of the oscillator with maximal amplitude. The frequency distribution of radiation is
\be
\frac{d N(x(s))}{dx(s)} = \frac{\pi \alpha q^2 s^2}{\beta^2_z (s) (1 + q^2 s^2)} x f(x(s)). 
\ee

The spectral line shape of the radiation from a channeled positron is formed at frequencies $x_{1,2}$, for which $Y(x(s)) = 0$, corresponding to the energy-momentum conservation law during radiation.

These frequencies are the roots of the equation
\be
\begin{gathered}
\varphi(x (s)) = x^2 - 2 x + \frac{\gamma_{\text{th}}}{\gamma} = 0, \\
\gamma_{\text{th}} (s) = \left(\frac{\omega_p}{\Omega_0}\right)^2 Q(s), \\
x_{1,2} (s) = 1 \mp \sqrt{1 - \frac{\gamma_{\text{th}}(s)}{\gamma}},
\end{gathered}
\ee
where $\gamma_{\text{th}} (s)$ is the threshold energy of a bunch required for radiation formation by positrons oscillating with amplitudes smaller than $s$ at frequencies $x_{1,2}$.

\section{Spectral line shape of a bunch at maximum threshold energy}
%%%%%%%%%%%%%%%%%%%%%%%%%%%%%%%%%%%%%%%%%%%%%%%%%%%%%%%%%%%%%%%%%%%%%%%%%%%%%%%%%%%%%%%%%%%%%%%%%%%%%%%%%%%%%%%%%%%%%%%%%%%%%%%%%%%%%%%%%
%%%%%%%%%%%%%%%%%%%%%%%%%%%%%%%%%%%%%%%%%%%%%%%%%%%%%%%%%%%%%%%%%%%%%%%%%%%%%%%%%%%%%%%%%%%%%%%%%%%%%%%%%%%%%%%%%%%%%%%%%%%%%%%%%%%%%%%%%

For the threshold energy $\gamma = \gamma_{\text{th}} (1)$, positrons with parameter $s=1$ radiate at frequency $x=1$, while for positrons with parameter $s$ ($0 < s < 1$) the radiation occurs at frequencies
\be
\begin{gathered}
x_{1,2} (s) = 1 \mp \frac{q}{\sqrt{1+ q^2}} \sqrt{1 - s^2}, \\
x_1 (s) \cdot x_2 (s) = \frac{\gamma_\text{th} (s)}{\gamma_\text{th} (1)}.
\end{gathered}
\ee

For large numbers of positron oscillations ($n \gg 1$), the line shapes of their radiation have a $\delta$-function–like form, with a peak height of $n^2$ and line width $\pi/n$. Then, to leading order in $1/n$, we have
\be
\begin{gathered}
f(x(s)) = \pi n \delta\left[Y(x(s)) \right] = \pi n \delta\left[\frac{\pi}{2} \frac{\varphi(x(s))}{x(s)} \right] = \\ 
= 2 n \beta_z (x) \delta\left[\Psi (x(s))\right], \\
\Psi (x(s)) = \frac{\varphi(x(s))}{x(s)} = x(s) - 2 + \frac{x_1(s) \cdot x_2(s)}{x(s)}.
\end{gathered}
\ee

If the argument of the $\delta$-function has two roots, the singularities can be regularized using
\be
\begin{gathered}
\delta \left[ \Psi(x(s)) \right] = \abs{\frac{x(s) - x_2 (s)}{\Psi^\prime (x(s))}}_{x(s) = x_1 (s)} \delta\left[ x(s) - x_1 (s) \right] +  \\ 
+ \abs{\frac{x(s) - x_1 (s)}{\Psi^\prime (x(s))}}_{x(s) = x_2 (s)} \delta\left[ x(s) - x_2 (s) \right] = \\ 
= x_2 (s) \delta\left[ x(s) - x_1 (s)\right] + x_1 (s) \delta\left[x(s) - x_2 (s) \right]. 
\end{gathered}
\ee

Consequently, the frequency distribution of radiation from a positron channeled in a nanotube is
\be
\begin{gathered}
\frac{d N[x(s)]}{d x(s)} = 2 \pi \alpha n q^2 F[x(s)], \\
F[x(s)] = \frac{s^2}{\beta_z (s) (1 + q^2 s^2)} \left( x(s) x_2 (s) \delta[x(s) - x_1 (s)] \right. \\
\left. + x(s) x_1 (s) \delta [x(s) - x_2 (s)]\right), \\
0 \leq s \leq 1 \quad 1 - \eta \leq x_1 (s) \leq 1; \quad 1 \geq s 
\geq 0 \quad 1 \leq x_2 (s) \leq 1 + \eta.
\end{gathered}
\ee

Using the relations
\be
\begin{gathered}
x_1 (s) + x_2 (s) = 2, \quad x_1 (s) \cdot x_2 (s) = 1 - \eta^2 (1 - s^2); \\
\beta_z (s) \approx 1 - \frac{1 + q^2 s^2}{2 \gamma^2 (1)} \quad \left(\gamma (1) \gg 1 \right); \\
s^2 = 1 - \frac{(1 - x_1 (s))^2}{\eta^2} = 1 - \frac{(x_2 (s) - 1)^2}{\eta^2}; \\
1 + q^2 s^2 = (1 + q^2) x_1 (s) (2 - x_1 (s)) = (1 + q^2) x_2 (s) (2 - x_1(s)) = \\
= (1 + q^2) x_1 (s) \cdot x_2 (s),
\end{gathered}
\ee
to leading order in $(1+q^2)/(2 \gamma^2(1))$ we obtain
\be
\begin{gathered}
F[x(s)] = \frac{1}{1+q^2} \left\{ \left(1 - \frac{(1 - x_1 (s))^2}{\eta^2} \right) + \left( 1 - \frac{(x_2 (s) - 1)^2}{\eta^2} \right) \right\} = \\
= \frac{1}{1 + q^2} \phi [x(s)].
\end{gathered}
\ee

%***************************************************************
\begin{figure}[t]
\bt{r}
\centerline{\includegraphics[width=1\columnwidth]{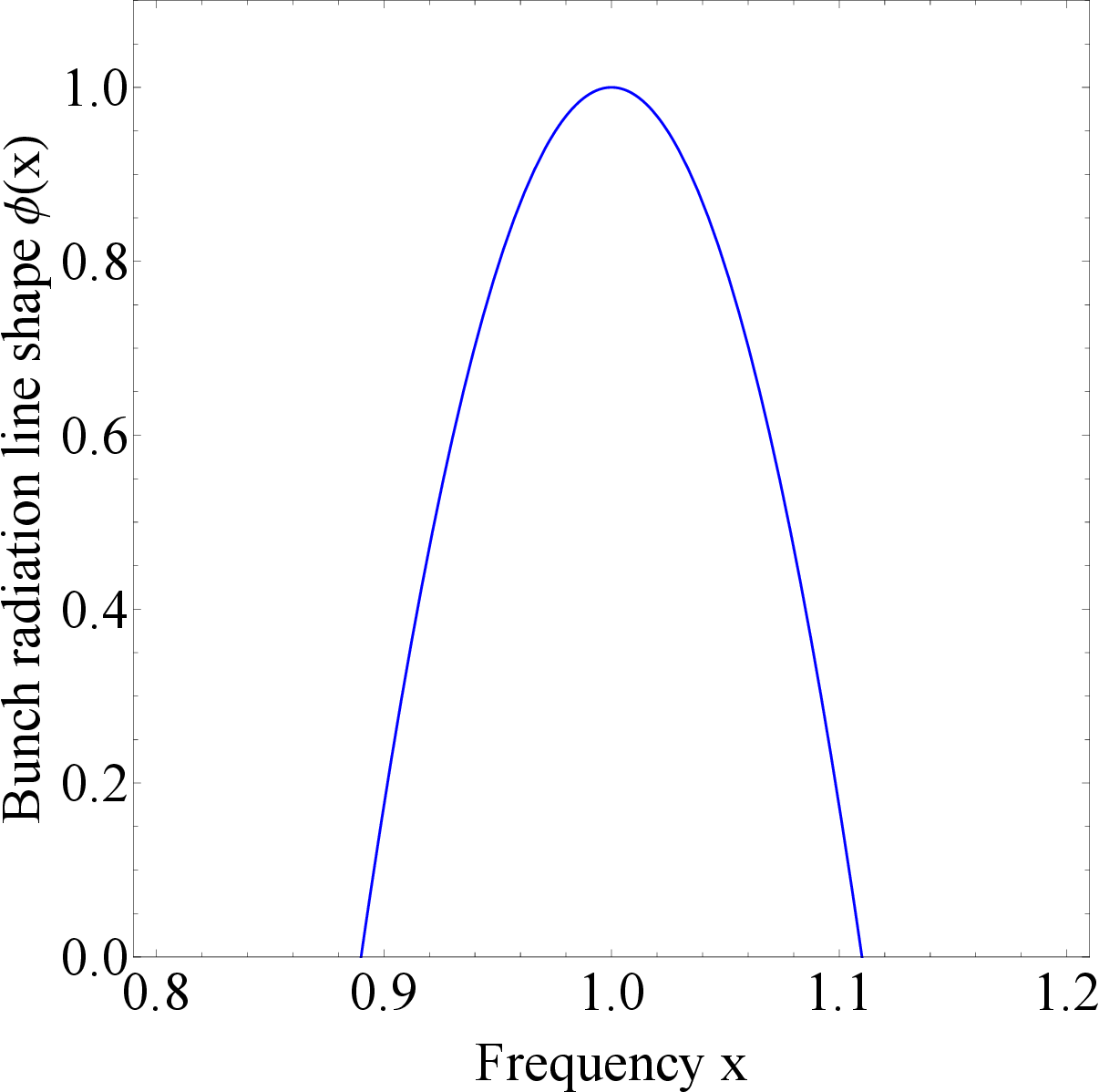}} 
\et
\caption{
Frequency distribution (spectral line shape) $\phi(x)$ of radiation from a bunch of positrons channeled in a nanotube, with the bunch energy equal to the maximum threshold energy.
}
\label{fig:BunchRadLineShape}
\end{figure}
%***************************************************************

The line shape of radiation from a bunch of channeled positrons is
\be
\begin{gathered}
\phi (x) = 1 - \frac{1 - x^2}{\eta^2}, \\
1 - \eta \leq x \leq 1 + \eta,
\end{gathered}
\ee
with a line width of order $\eta = q / \sqrt{1 + q^2}$.

The number of photons radiated by a bunch of $N_b$ channeled positrons is
\be
\begin{gathered}
N_{\text{ph}} (x) = 2 \pi \alpha n \eta^2 N_b \int\limits_{1-\eta}^{1+\eta} \left( 1 - \frac{(x-1)^2}{\eta^2} \right) \, dx = \frac{8 \pi \alpha n \eta^3}{3} N_b. 
\end{gathered}
\ee

% \be
% \begin{gathered}
% N_1 [x(s)] = \frac{2 \pi \alpha n q^2}{1 + q^2} \left\{ \int\limits_{1 - \eta}^{1} \phi [x_1 (s)] \, d x_1 (s) + \int\limits_{1}^{1 + \eta} \phi [x_2 (s)] \, d x_2 (s)\right\} = \\ 
% = 2 \pi \alpha n \eta^2 \left(\frac23 \eta + \frac23 \eta \right) = \frac{8 \pi \alpha n}{3} \eta^3. 
% \end{gathered}
% \ee

Thus, a positron bunch of $N_b$ particles channeled in a nanotube with energy $\gamma = \gamma (1)$ and sufficiently small energy and angular divergence will emit $N_\text{ph}$ photons with energy $\hbar \omega = \hbar \Omega_0 \gamma^{3/2} (1)$ and line width $\eta$.

\subsection{Example}

Consider a carbon nanotube with radius $R = 7$~\SI{}{\angstrom} and potential barrier $U_0 = 32$~\SI{}{\electronvolt} ($\nu = 6.26 \cdot 10^{-5}$), in a dispersive medium with plasma energy $\hbar \omega_p = 31$~\SI{}{\electronvolt}. For a positron bunch with energy $\gamma = \gamma(1) = 96.647$ ($E = 49.4$~\SI{}{\mega\electronvolt}), we have $\sqrt{2\nu} = 11.19 \cdot 10^{-3}$, $\hbar \Omega_0 = 3.155$~\SI{}{\electronvolt} ($l_0 = 3.93 \cdot 10^{-5}$~\SI{}{\centi\meter}), $q = \beta_\perp \gamma = \sqrt{2 \nu \gamma} = 0.11$ ($\eta = q/\sqrt{1+q^2} = 0.1095$, $\eta^3 = 1.3 \cdot 10^{-3}$).

The spatial period of oscillation is $l = l_0 \sqrt{\gamma} = 3.864 \cdot 10^{-4}$~\SI{}{\centi\meter}, the total nanotube length $L = n l = 1$~\SI{}{\centi\meter}, $n = 2.588 \cdot 10^3$. The number of photons radiated with energy $3$~\SI{}{\kilo\electronvolt} ($4.1$~\SI{}{\angstrom}) is $0.2077 N_b$.

\section{Conclusion}
%%%%%%%%%%%%%%%%%%%%%%%%%%%%%%%%%%%%%%%%%%%%%%%%%%%%%%%%%%%%%%%%%%%%%%%%%%%%%%%%%%%%%%%%%%%%%%%%%%%%%%%%%%%%%%%%%%%%%%%%%%%%%%%%%%%%%%%%%
%%%%%%%%%%%%%%%%%%%%%%%%%%%%%%%%%%%%%%%%%%%%%%%%%%%%%%%%%%%%%%%%%%%%%%%%%%%%%%%%%%%%%%%%%%%%%%%%%%%%%%%%%%%%%%%%%%%%%%%%%%%%%%%%%%%%%%%%%

%сильноточный позитронный сгусток

The problem of the zero-angle radiation characteristics of a bunch of positrons channeled in a nanotube, with energy equal to the maximum value of the amplitude-dependent threshold energy for radiation formation, is considered. With decreasing oscillation amplitude, symmetric frequencies relative to the fundamental frequency are generated, the spacing of which increases as the amplitude decreases. This leads to the formation of the bunch radiation line shape. Taking this effect into account, the frequency distribution of the bunch radiation has been obtained. The number of radiated photons and the line width of the bunch radiation have been determined. The formation of radiation in a system of carbon nanotubes, considering the medium polarization and using a medium-energy positron bunch, is also numerically analyzed. Using a high-current positron bunch with an energy of 50~\si{\mega\electronvolt}, one can generate an intense, directed, quasi-monochromatic soft X-ray beam with a relative line width of approximately 0.1 and circular polarization. Such photon beams have significant scientific and practical applications.

% The problem about the channeling radiation of the positron bunch in nanotube is solved in the dipole approximation. The dipole approximation legitimately is applied for the large values of the oscillation parameter also, if the emission occurs at very small angles. 

% The main contribution into the radiation makes the first harmonic of radiation which is formed at a zero angle too that it is important from the practical point of view. The soft photons are emitted at a zero angle also because of the medium polarization. The numbers of soft and hard photons are of the same order of magnitude. There are two energy thresholds, namely, the lower threshold of the medium polarization and the upper threshold for positrons oscillating in the channel with the maximum amplitude: the higher the initial radial coordinate of a positron, the greater the threshold energy. If the bunch energy is equal to the maximum value of the threshold energy, then the all channeled positrons contribute to the radiation in nanotube.

% Hence, the monochromatic beam of high energy photons with circular polarization is formed in nanotube due to the cylindrical symmetry. The radiation beam of soft X-ray photons has important practical significance.


%\appendix

%\section{Sample Appendix Section}

 \bibliographystyle{elsarticle-num} 


\end{document}